\documentstyle[12pt]{article}

\author{Yu. G. Ignat'ev \& A. A. Popov\thanks{E-mail: popov@kspu.ksu.ras.ru}\\
Department of Geometry, Kazan State Pedagogical University,\\
Mezhlauk 1, Kazan 420021, Russia}
\title{Spherically symmetric perturbation of ultrarelativistic fluid in\\
homogeneous and isotropic universe}
\date{}
\begin{document}

\maketitle
\begin{abstract}
A solution of the linearized Einstein's equations for a spherically
symmetric per\-tur\-ba\-tion of the ultrarelativistic fluid in the
homogeneous and isotropic universe is obtained. Conditions on the
boundary of the perturbation are discussed. The examples of
particle-like and wave-like solutions are given.
\end{abstract}

\section{Introduction}

The description of metric perturbations in relativistic cosmology is
significance in the theory of creation of the universe
large-scale structure. Spherically symmetric perturbations are
an important and interesting class of that perturbations.

It is easy to understand that the spherically symmetric perturbation may be
represent as outgoing and ingoing waves (traveling from and into the center
of the configuration). The ingoing wave amplitude increases as soon as the
wave approaches to the center of the configuration. Therefore the nonlinear
stage of the development of the spherically symmetric perturbation must
appear more quickly in the contrast to the case considered by Lifschitz
\cite{1} for flatwave perturbations. As  the wave velosity in the
ultrarelativistic fluid is equal to the velosity of sound $c/\sqrt{3}$ so the
time $T$ of appearing of the nonlinear stage is only determined by the
typical size $L$ of an initial perturbation: $T=\sqrt{3}Lc^{-1}.$
Below we describe this process explicitly.

\section{Ultrarelativistic fluid in spherically symmetric space-time.
Rigorous equations}

The spherically symmetric metric may be taken in the form \cite{2}
\begin{equation}
\label{1.1}ds^2=e^\nu d\eta ^2-e^\lambda \left[ dr^2+r^2\left( d\theta
^2+\sin {}^2\theta \ d\varphi ^2\right) \right] ,
\end{equation}
where the metric parameters $\nu $ and $\lambda $ are the functions of the
time coordinate $\eta $ and a radial space coordinate $r$ . With this
notations the nontrivial Einstein's equations take the form ($c=G=1,
\varepsilon=3p$)
\begin{equation}
\label{1.2}\frac{e^{-\lambda }}2\left[ \frac 12\lambda ^{\prime \ 2}+\lambda
^{\prime }\nu ^{\prime }+\frac 2r\left( \lambda ^{\prime }+\nu ^{\prime
}\right) \right] -\frac{e^{-\nu }}3\left( \stackrel{..}{\lambda }-\frac 12
\stackrel{.}{\lambda }\stackrel{.}{\nu }+\frac 34\stackrel{.}{\lambda }
^2\right) =8\pi p\left( 1+4\upsilon ^2\right) ,
\end{equation}
\begin{equation}
\label{1.3}\frac{e^{-\lambda }}4\left[ 2\left( \lambda ^{\prime \prime }+\nu
^{\prime \prime }\right) +\nu ^{\prime \ 2}+\frac 2r\left( \lambda ^{\prime
}+\nu ^{\prime }\right) \right] -\frac{e^{-\nu }}3\left( \stackrel{..}{
\lambda }-\frac 12\stackrel{.}{\lambda }\stackrel{.}{\nu }+\frac 34\stackrel{
.}{\lambda }^2\right) =8\pi p,
\end{equation}
\begin{equation}
\label{1.4}-e^{-\lambda }\left( \lambda ^{\prime \prime }+\frac 14\lambda
^{\prime \ 2}+\frac 2r\lambda ^{\prime }\right) +\frac{e^{-\nu }}4\stackrel{.
}{\lambda }^2=8\pi p\left( 3+4\upsilon ^2\right) ,
\end{equation}
\begin{equation}
\label{1.5}\frac{e^{-\lambda }}{2\sqrt{3}}\left( 2\stackrel{.}{\lambda }
^{\prime }-\nu ^{\prime }\stackrel{.}{\lambda }\right) =8\pi e^{\left( \nu
-\lambda \right) /2}4p\upsilon \sqrt{1+\upsilon ^2},
\end{equation}
where $\varepsilon$ is the energy density and $p$ is the pressure; $\upsilon
=u^re^{\lambda /2}$ is the frame projection of radial velisity and $u^r $ is
the radial coordinate of 4-velosity of fluid; a dot and a prime denote partial
derivatives with respect to
\begin{equation}
\label{1.6}\tau =\frac {\eta} {\sqrt{3}}
\end{equation}
and $r$, respectively.

There are two consequences of these equations
\begin{equation}
\label{1.7}\frac 12e^{-\lambda }\left[ -\lambda ^{\prime \prime }-\nu
^{\prime \prime }+\frac 12\lambda ^{\prime \ 2}+\lambda ^{\prime }\nu
^{\prime }-\frac 12\nu ^{\prime \ 2}+\frac 1r\left( \lambda ^{\prime }+\nu
^{\prime }\right) \right] =8\pi \upsilon ^24p,
\end{equation}
\begin{eqnarray}
\label{1.8}
-e^{-\lambda }\left[ 2\lambda ^{\prime \prime }+\nu ^{\prime \prime }+\frac
12\left( \lambda ^{\prime \ 2}+\lambda ^{\prime }\nu ^{\prime }+\nu ^{\prime
2}\right) +\frac 2r\left( 2\lambda ^{\prime }+\nu ^{\prime }\right) \right]
\nonumber \\
+e^{-\nu }\left( \stackrel{..}{\lambda }+\stackrel{.}{\lambda }^2-\frac 12
\stackrel{.}{\lambda }\stackrel{.}{\nu }\right) =0.
\end{eqnarray}

\section{Background spacetime}

In isotropic spherical coordinates the metric of the background space-time is
\begin{equation}
\label{2.1}ds^2=a^2\left\{ d\eta ^2-\frac 1{y^4}\left[ dr^2+r^2\left(
d\theta ^2+\sin {}^2\theta \ d\varphi ^2\right) \right] \right\},
\end{equation}
where
\begin{equation}
\label{2.2}y=\sqrt{1+Kr^2 b^{-2}},
\end{equation}
\begin{equation}
\label{2.3}a=\left\{
\begin{array}{ll}
a_0\tau , & \hspace{2cm} \mbox{for}\ K=0; \\ a_0\sinh (\sqrt{12}\tau
b^{-1}), & \hspace{2cm} \mbox{for}\ K=-1; \\ a_0\sin (\sqrt{12}\tau b^{-1}),
& \hspace{2cm} \mbox{for}\ K=+1,
\end{array}
\right.
\end{equation}
$K=0,$ $\pm 1,$ $b,a_0$ are the constant parameters of the background
space-time.

The energy density $\varepsilon_0$, the pressure $p_0$ and the frame
projection of radial velosity of the backgound fluid $\upsilon _0$ are
\begin{equation}
\label{2.4}\varepsilon_0=3p_0=\frac 1{8\pi }\left[ 4\ \frac{y^3}{a^2}\left(
\ y^{\prime \prime }-\frac 2y\ y^{\prime \ 2}+\frac 2r\ y^{\prime }\ \right)
+\frac 9{a^4}\stackrel{.}{a}^2\right] ,
\end{equation}
\begin{equation}
\label{2.5}\upsilon _0=0.
\end{equation}

\section{Perturbation equations and general solution}

Let us define small perturbation quantities as follows
\begin{equation}
\label{3.1}\delta \varepsilon=3\delta p=\varepsilon-\varepsilon_0,
\end{equation}
\begin{equation}
\label{3.2}\delta \nu=\nu-\ln \left( a^2\right),
\end{equation}
\begin{equation}
\label{3.3}\delta \lambda =\lambda -\ln \left( a^2/y^4\right) .
\end{equation}
We will assume that the frame projection of radial velosity of fluid $
\upsilon $ is the small quantity too. Substituting relations (\ref{3.1}-\ref
{3.3}) into Eq.(\ref{1.7}) and keeping only the first-order quantities in
perturba\-tion we obtain
\begin{equation}
\label{3.4}-\left( \delta \lambda ^{\prime \prime }+\delta \nu ^{\prime
\prime }\right) +\frac 1r\left( \delta \lambda ^{\prime }+\delta \nu
^{\prime }\right) -\frac{4y^{\prime }}y\left( \delta \lambda ^{\prime
}+\delta \nu ^{\prime }\right) =0.
\end{equation}
The solution of this equation is
\begin{equation}
\label{3.5}\delta \lambda +\delta \nu =\left\{
\begin{array}{ll}
C_1+C_2r^2, & \hspace{2cm}\mbox{for}\ K=0; \\ C_3+C_4b^2K^{-1}y^{-2}, &
\hspace{2cm}\mbox{for}\ K=\pm 1,
\end{array}
\right.
\end{equation}
where $C_1,C_2,C_3\ \mbox{and} \ C_4$ are the arbitrary functions of the
time coordinate $\tau $.

If we require that
\begin{equation}
\label{3.6}\delta \lambda = \delta \nu = 0
\quad \mbox{for} \ r > r_0 (\tau) ,
\end{equation}
where $r_0$ is the some function of $\tau$,
i.e. the space-time is asymptotically homogeneous and isotropic, then
\begin{equation}
\label{3.7}C_1=C_2=C_3=C_4=0
\end{equation}
and
\begin{equation}
\label{3.8}\delta \lambda =-\delta \nu .
\end{equation}
Substituting the Eqs.(\ref{3.2}), (\ref{3.3}) and (\ref{3.8}) into Eq.(\ref
{1.8}), we get in the linear approximation
\begin{equation}
\label{3.9}-y^4\left[ \delta \lambda ^{\prime \prime }+2(\frac 1r-\frac{
y^{\prime }}y)\delta \lambda ^{\prime }\right] +\delta \ddot \lambda
+4\frac{\dot a}a\delta \dot \lambda +4\frac{\ddot a}a \delta \lambda =0.
\end{equation}
If we introduce new radial coordinate
\begin{equation}
\label{3.10}L(r)=\int \limits_0^r \frac{dr}{1+K(r/b)^2}=\left\{
\begin{array}{ll}
r, & \hspace{2cm}K=0; \\ b\mathop{\rm Artanh}(rb^{-1}), & \hspace{2cm}K=-1; \\
b\arctan (rb^{-1}), & \hspace{2cm}K=+1
\end{array}
\right.
\end{equation}
and make the substitution
\begin{equation}
\label{3.11}\delta \lambda =\frac{y^2}{ar}\frac \partial {\partial \tau
}\left( \frac \varphi a\right) ,
\end{equation}
where $\varphi =\varphi (\tau ,L)$, then the equation (\ref{3.9}) can be
rewritten in the form
\begin{equation}
\label{3.12}\frac \partial {\partial \tau }\left[ \frac 1a\left( \stackrel{..
}{\varphi }-\frac{\partial ^2\varphi }{\partial L^2}-\frac{4K}{b^2}\varphi
\right) \right] =0.
\end{equation}
One can easily integrate this equation
\begin{equation}
\label{3.13}\stackrel{..}{\varphi }-\frac{\partial ^2\varphi }{\partial L^2}-
\frac{4K}{b^2}\varphi =aF\left( L\right) ,
\end{equation}
where $F\left( L\right) $ is an arbitrary function of $L$.

A general solution of Eq.(\ref{3.13}) is the sum of the general solution of
the equation
\begin{equation}
\label{3.14}\stackrel{..}{\varphi }-\frac{\partial ^2\varphi }{\partial L^2}-
\frac{4K}{b^2}\varphi =0
\end{equation}
and the particular solution of the equation (\ref{3.13}). But this
particular solution does not change the form $\delta \lambda $, because it
can be rewritten as $aW$, where $W=W(L)$ and
\begin{equation}
\label{3.13*}-\frac{d^2W}{dL^2}-\frac{16K}{b^2}W=F.
\end{equation}
Therefore we can substitute the solution of equation (\ref{3.14}) but not
equation (\ref{3.13}) into the expression (\ref{3.11}). In the case $K=0$
this solution is
\begin{equation}
\label{3.15}\varphi (\tau ,L)=\Phi _{+}(\tau +L)+\Phi _{-}(\tau -L)
\end{equation}
and in the case $K=\pm 1$
\begin{eqnarray}
\varphi (\tau ,L)=\int \limits_0^{\tau +L}\frac{d^3f\left( t\right) }{dt^3}
J_0\left( \frac{2i}b\sqrt{\left( \tau -L\right) \left( \tau
+L-t\right) }\right) dt \nonumber\\
+\int\limits_0^{\tau -L}\frac{d^3g\left(
t\right) }{dt^3} J_0\left(
\frac{2i}b\sqrt{\left( \tau +L\right) \left( \tau -L-t\right) }\right) dt
\nonumber\\
\label{3.16}
+b\left[ \frac{d^3}{dt^3}\left( f+g\right) \right] _{\mid
t=0} J_0\left( 2i\frac{\tau ^2-L^2}{b^2}\right) ,
\end{eqnarray}
where $\Phi _{+},\ \Phi _{-},\ f\ \mbox{and} \ \ g$ are arbitrary functions
and $J_0$ is the Bessel function.

Thus the relations (\ref{3.11}) and (\ref{3.8}) describe the
spherically symmetric perturbation of the metric satisfying the
condition (\ref{3.6}) on the background of spacetime (\ref{2.1}), if we
take into account the relations (\ref{3.15}) and (\ref{3.16}).

The perturbations of the density $\delta \varepsilon$ and the frame
projection of the radial velosity $\upsilon $ are
\begin{eqnarray}
8\pi \delta \varepsilon=8\pi
3\delta p=\frac{\stackrel{.}{a}}{a^3}\left(
\delta \stackrel{.}{\lambda }+\frac{\stackrel{.}{a}}a\delta \lambda \right)
- \frac{y^4}{a^2}\left[ \delta \lambda ^{\prime \prime }+2\left( \frac
1r-\frac{y^{\prime }}y\right) \delta \lambda ^{\prime } \right. \nonumber \\
\label{3.17}\left. +\frac 4y\left( y^{\prime \prime }-2\frac{y^{\prime \ 2}}
y+\frac 2ry^{\prime }\right) \delta \lambda \right] \ ,
\end{eqnarray}
\begin{equation}
\label{3.18}8\pi \upsilon =\frac{\sqrt{3}y^2}{4\varepsilon
_0}\left( \delta
\stackrel{.}{\lambda }^{\prime }+\frac{\stackrel{.}{a}}a\delta \lambda
^{\prime}\right) .
\end{equation}

In the case
\begin{equation}
\label{3.19}K=0,\ \Phi _{+}\left( x\right) =D_{+}+D_1\ x+D_2\ x^2+D_3\ x^3,\
\Phi _{-}\left( x\right) =D_{-}+D_1\ x-D_2\ x^2+D_3\ x^3,
\end{equation}
where $D_{+},\ D_{-},\ D_1,\ D_2,\ D_3$ are constants, we obtain a Newtonian
potential \cite{3}
\begin{equation}
\label{3.20}\delta \lambda =-\ \delta \nu =\frac {2m}{ar}
\end{equation}
caused by a particle of variable mass
\begin{equation}
\label{3.21}m=-\ \frac{\stackrel{.}{a}}{2a^2}\left( D_{+}+D_{-}\right) +
\frac{\tau ^2}a2D_3\ .
\end{equation}

In the conclusion of this section we would like to note that the
perturbations in the homogeneous and isotropic universe are gauge-dependent
\cite{6}. Therefore, we should know whether the per\-turbative quantities
$\delta \varepsilon, v, \delta \lambda, \delta \nu$ obtained in this
section are the actual physical perturbations or merely an artifact
of gauge.

In concequence of the symmetry of problem the perturbations described
in this paper are scalar ones.
The gauge-invariant amplitude of density perturbation with the notations of
the paper \cite{6} is
\begin{equation}
\label{3.22}\epsilon_g=\delta - 3(1+ \mathop{\rm w}) \frac 1{\mathop{\rm k}}
\frac{\dot S}S \left( B^{(0)} - \frac 1{\mathop{\rm k}} \dot H^{(0)}_T
\right)
\end{equation}
The choice of the isotropic coordinates (\ref{1.1}) in our paper corresponds
to a longitudinal gauge. With the notations of the paper \cite {6} this give
\begin{equation}
\label{3.23}B^{(0)} = H^{(0)}_T = 0.
\end{equation}
By comparing the corresponding expressions in two papers we find
\begin{equation}
\label{3.24} \epsilon_g Q^{(0)}=\delta Q^{(0)}=\delta \varepsilon /
\varepsilon_0 .
\end{equation}
The relations between the other perturbative quantities can be determined
analogously
\begin{equation}
\label{3.25}v^{(0)}_s Q^{(0)}=\left[ v^{(0)} - \frac1{\mathop{\rm k}}
\dot H^{(0)}_T \right] Q^{(0)}=v^{(0)} Q^{(0)}=v,
\end{equation}
\begin{equation}
\label{3.26} \Phi_A Q^{(0)}=\left[ A+\frac 1{\mathop{\rm k}}
\dot B^{(0)} + \frac 1{\mathop{\rm k}} \frac {\dot S}S B^{(0)} -
\frac 1{{\mathop{\rm k}}^2} \left( \ddot H^{(0)}_T +
\frac{\dot S}S \dot H^{(0)}_T \right) \right] Q^{(0)}=A Q^{(0)}=
\frac{\delta \nu}2,
\end{equation}
\begin{equation}
\label{3.27} \Phi_H Q^{(0)}=\left[ H_L + \frac13 H^{(0)} + \frac1
{\mathop{\rm k}} \frac{\dot S}S B^{(0)} - \frac 1{\mathop{\rm k}^2}
\frac{\dot S}S \dot H^{(0)}_T \right] Q^{(0)}=H_L Q^{(0)}=
\frac{\delta \lambda}2.
\end{equation}
>From this relations we find that the perturbation quantities introduced
in this section are actualy physical as well as the gauge-invariant
quantities $\epsilon_g Q^{(0)}, v^{(0)}_s Q^{(0)}, \Phi_A Q^{(0)},
\Phi_H Q^{(0)}$.

\section{Boundary conditions}

The solution (\ref{3.20}) is physically unacceptable for the description of
the gravitational field of perturbation which arose as a result of
fluctuation at the moment of ''time'' $\tau =\tau _o$, since it is
inconsistent with the principle of causality. Therefore, the boundary
conditions should be formulated in the way that at least beyond the light
horizon, the potential $\delta \lambda $ should vanish together with its
derivatives. Such boundary conditions are in according with the ''birth'' of
a perturbation as a result of the redistribution of Robertson-Walker matter.
In fact, however, the horizon of the perturbation is not the light cone but
the sound cone, since the density perturbations extend with
velosity of sound. Thus the boundary conditions at the sound horizon $L=\tau
-\tau _0$ are taken in the form
\begin{equation}
\label{4.1}\delta \lambda _{\mid L=\tau -\tau _0}=\left. \frac
\partial {\partial L}\delta \lambda \right| _{L=\tau -\tau
_0}=\left. \frac \partial {\partial \tau} \delta \lambda
\right| _{L=\tau -\tau _0}=0.
\end{equation}
Substituting Eqs.(\ref{3.15}) and (\ref{3.16}) into this relations, we
obtain in the case $K=0$
\begin{equation}
\label{4.2}\Phi _{+}^{\prime }\left( 2\tau -\tau _0\right) +\Phi
_{-}^{\prime }\left( 0\right) -\frac 1\tau \left[ \Phi _{+}\left( 2\tau
-\tau _0\right) +\Phi _{-}\left( 0\right) \right] =0,
\end{equation}
\begin{equation}
\label{4.3}\Phi _{+}^{\prime \prime }\left( 2\tau -\tau _0\right) -\Phi
_{-}^{\prime \prime }\left( 0\right) -\frac 2\tau \Phi _{+}^{\prime }\left(
2\tau -\tau _0\right) +\frac 1{\tau ^2}\left[ \Phi _{+}\left( 2\tau -\tau
_0\right) +\Phi _{-}\left( 0\right) \right] =0,
\end{equation}
\begin{equation}
\label{4.4}\Phi _{+}^{\prime \prime }\left( 2\tau -\tau _0\right) +\Phi
_{-}^{\prime \prime }\left( 0\right) -\frac 3\tau \left[ \Phi _{+}^{\prime
}\left( 2\tau -\tau _0\right) +\Phi _{-}^{\prime }\left( 0\right) \right]
+\frac 3{\tau ^2}\left[ \Phi _{+}\left( 2\tau -\tau _0\right) +\Phi
_{-}\left( 0\right) \right] =0,
\end{equation}
and in the case $K=\pm 1$
\begin{eqnarray}
f^{\prime \prime \prime }\left( 2\tau -\tau _0\right) +g^{\prime \prime
\prime }\left( 0\right) -\frac{\stackrel{.}{a}}a\left\{ f^{\prime \prime
}\left( 2\tau -\tau _0\right) -f^{\prime \prime }\left( 0\right)
+b\left[ f^{\prime \prime \prime }\left( 0\right) +g^{\prime \prime
\prime }\left(
0\right) \right] \right\} \nonumber\\
\label{4.5}-\frac{2\tau }{b^2}f^{\prime \prime }\left( 0\right) +\frac{
f^{\prime }\left( 2\tau -\tau _0\right) -\ f^{\prime }\left( 0\right)
}{b^2} =0,
\end{eqnarray}
\begin{eqnarray}
f^{\prime \prime \prime \prime }\left( 2\tau -\tau _0\right)
-g^{\prime
\prime \prime \prime }\left( 0\right) -\frac{\stackrel{.}{a}}a\left[
f^{\prime \prime \prime }\left( 2\tau -\tau _0\right) -g^{\prime \prime
\prime }\left( 0\right) \right]-\frac{8\tau ^2}{b^3}\left[ f^{\prime
\prime \prime }\left(0\right) +g^{\prime \prime \prime }\left( 0\right)
\right] \nonumber\\
 -\frac{2\tau}{b^2} g^{\prime \prime \prime }\left( 0\right) +\left(
\frac{2\tau^2}{b^4}-\frac{\stackrel{.}{a}}a\frac{2\tau }{b^2}\right)
f^{\prime \prime }\left( 0\right)
+\frac{\stackrel{.}{a}}{ab^2}f^{\prime }\left( 2\tau -\tau _0\right)
+\left( \frac{2\tau }{b^4}-\frac{\stackrel{.}{a}}{ab^2}\right)
f^{\prime }\left( 0\right)\nonumber\\
\label{4.6} -\frac 1{b^4}\left[ f\left( 2\tau -\tau
_0\right) -f\left( 0\right) \right] =0,
\end{eqnarray}
\begin{eqnarray}
f^{\prime \prime \prime \prime }\left( 2\tau -\tau _0\right)
+g^{\prime
\prime \prime \prime }\left( 0\right) +\left( \frac{8\tau ^2}{b^3}+\frac{12k}
b\right) \left[ f^{\prime \prime \prime }\left( 0\right) +g^{\prime
\prime
\prime }\left( 0\right) \right] +\frac{2\tau }{b^2}g^{\prime \prime \prime
}\left( 0\right)\nonumber\\
+\frac 2{b^2}\left( 1+6
k\right) \left[ f^{\prime \prime }\left(
2\tau -\tau _0\right) -f^{\prime \prime }\left( 0\right) \right]
-\frac{ 2\tau ^2}{b^4}f^{\prime \prime }\left( 0\right) -\frac{2\tau
}{b^4}f^{\prime }\left( 0\right)\nonumber\\
\label{4.7}+\frac 1{b^4}\left[ f\left( 2\tau
-\tau _0\right) -f\left( 0\right) \right] =0,
\end{eqnarray}
where a prime denotes a derivative of the function with respect to its
argument.

\section{Case of spatially flat universe ($K=0$)}

Integrating Eqs.(\ref{4.2})-(\ref{4.4}), we obtain
\begin{equation}
\label{5.1}\Phi _{+}\left( x\right) =-\frac{x^2}2\Phi _{-}^{\prime \prime
}\left( \tau _0\right) -x\left[ \Phi _{-}^{\prime }\left( \tau _0\right)
-\tau _0\Phi _{-}^{\prime \prime }\left( \tau _0\right) \right] -\frac{\tau
_0^2}2\Phi _{-}^{\prime \prime }\left( \tau _0\right) +\tau _0\Phi
_{-}^{\prime }\left( \tau _0\right) -\Phi _{-}\left( \tau _0\right) .
\end{equation}
We denote a new function
\begin{eqnarray}
\Psi \left( \tau -r-\tau _0\right) =\Phi _{-}\left( \tau -r\right) -\Phi
_{-}\left( \tau _0\right) -\frac 1{1!}\Phi _{-}^{\prime }\left( \tau
_0\right) \cdot \left( \tau -r-\tau _0\right) \nonumber\\
\label{5.2}-\frac 1{2!}\Phi _{-}^{\prime \prime }\left( \tau _0\right) \cdot
\left( \tau -r-\tau _0\right) ^2.
\end{eqnarray}
Then it is possible to reduce the relation (\ref{3.11}) to
\begin{equation}
\label{5.3}\delta \lambda =\frac 1{ar}\frac \partial {\partial \tau }\left[
\frac{\Psi \left( \tau -r-\tau _0\right) }a\right]
\end{equation}
and the conditions (\ref{4.2})-(\ref{4.4}) to
\begin{equation}
\label{5.4}\Psi \left( 0\right) =\Psi ^{\prime }\left( 0\right) =\Psi
^{\prime \prime }\left( 0\right) =0.
\end{equation}
If the function $\Psi \left( x\right) $ satisfies to the condition
\begin{equation}
\label{5.5}\Psi \left( x\right) =0\qquad \mbox{for}\ x<0,
\end{equation}
then the relations (\ref{5.3})-(\ref{5.4}) describe the spherically
symmetric fluctuation of gravitational field in region $\tau -r-\tau _0>0$
that satisfies to the conditions
\begin{equation}
\label{5.6}\left. \delta \lambda \right| _{r=\tau -\tau _0}=\left. \frac
\partial {\partial r}\delta \lambda \right| _{r=\tau -\tau _0}=\left.
\frac \partial {\partial \tau}\delta {\lambda }\right| _{r=\tau -\tau _0}=0.
\end{equation}

For more detailed study of the solution near the boundary we represent the
function $\Psi \left( x\right) $ as a power series
\begin{equation}
\label{5.7}\Psi \left( x\right) =\sum\limits_{n=0}^\infty \Psi _n\ x^n H(x),
\end{equation}
where $\Psi _k$ are constants and
\begin{equation}
\label{5.71}H(x)=\left\{
\begin{array}{ll}
0, & x<0, \\
1, & x \geq 0.
\end{array}
\right.
\end{equation}
Taking the relation (\ref{5.4}) into account, we obtain
\begin{equation}
\label{5.8}\Psi _0=\Psi _1=\Psi _2=0.
\end{equation}
Then relation (\ref{5.3}) reduces to
\begin{equation}
\label{5.9}\delta \lambda =\frac 1{ar}\sum\limits_{n=3}^\infty \frac{\Psi _n}
a\left( \tau -r-\tau _0\right) ^{n-1}\left( n-1+\frac{r+\tau _0}\tau
\right) H(\tau -r-\tau _0).
\end{equation}
The solution obtained in \cite{4}
\begin{equation}
\label{5.10}\delta \lambda =\Psi _3\frac{2\tau ^2}a\left( \frac 1{ar}-\frac
3{2a\tau }+\frac{r^2}{2a\tau ^3}\right) H(\tau -r)
\end{equation}
corresponds to the first term of the series (\ref{5.9}) for $\tau _0=0$.
The solution (\ref{5.3}) is a particle-like one if the function
\begin{equation}
m=\frac \partial {\partial \tau }\left[ \frac{\Psi \left( \tau -\tau
_0\right) }{2a}\right]
\end{equation}
is not equal to zero. This function describes the mass of particle in the
center of configuration and in the case considered in \cite{4} it is
\begin{equation}
m=\frac{\Psi _3\tau ^2}a.
\end{equation}

The wave traveling from the center of configuration
\begin{equation}
\label{5.11}\delta \lambda =\left\{
\begin{array}{ll}
\displaystyle{\frac{\Psi _0}{ar\tau }}\left[ \displaystyle{\frac{4\pi }l}
\sin \displaystyle{\frac{2\pi }lx} +\displaystyle{\frac 1\tau} \cos
\displaystyle{\frac{2\pi }lx} -\displaystyle{\frac 1\tau} \right] \left[
1-\cos\displaystyle{\ \frac{2\pi }lx} \right], & \mbox{for} \ 0\leq x \leq
l ; \\ 0, & \mbox{for} \ x<0, x>l,
\end{array}
\right.
\end{equation}
where $x = \tau -r-\tau _0, \Psi _0$ and $l$ are constants, may be an
example of nonparticle-like solution in the case $\tau>\tau_0+l$.
The function $\Psi $ that corresponds
to that solution is
\begin{equation}
\label{5.12}\Psi(x) =\left\{
\begin{array}{ll}
\Psi _0\left[ 1-\cos\displaystyle{\frac{2\pi}{l}x} \right] ^2, & \mbox{for}
\quad 0\leq x \leq l; \\ 0, & \mbox{for} \quad x<0, x>l.
\end{array}
\right.
\end{equation}

\section{Conclusion}

We have considered the spherically symmetric perturbation of the
cosmological perfect fluid with the equation of state $p=\varepsilon /3$.
The general solution of the linearized Einstein's equations which describe
this perturbation has been obtained. The solution is represented as
outgoing and ingoing waves moving with velosity of sound.
In order to consider the perturbation in causally connected region we
have imposed the boundary conditions on the solution.
In this case we have investigated the behavior of solution near
the boundary which is the sound horizon. We have given the
examples of particle-like and wave-like solutions.

We would like to finish our article by the following remark. The point of
view that the large-scale universe structure is pancake-like is well-known
\cite{5}. This point of view is based on the fact that flat perturbations
(with the wave length more than the Jeans one) begin to be increasing only
with the decoupling moment on the nonrelativistic
stage of the universe extension. However, the obtained solution shows
that spherical perturbations increase already on the ultrarelativistic
universe extension stage. Thus it may be turned out that at the decoupling
moment the spherical perturbations will have a bigger amplitude than the
flat ones, and in this case we will have to revise the scenario of
creation of large-scale universe structure.

\end{document}